\documentstyle[rmaaconf]{article}
\begin{document}

\title{THE STATE OF THE GAS AROUND YOUNG STELLAR GROUPS}

\author{Jos\'e Franco\altaffilmark{1},
Tomasz Plewa\altaffilmark{2}
\& Guillermo Garc\'\i a-Segura\altaffilmark{1}}
\altaffiltext{1}{Instituto de Astronom\' \i a--UNAM, Apdo. Postal 70-264, 
04510 M\'exico D. F., M\'exico}
\altaffiltext{1}{Max-Planck-Institut f\"ur Astrophysik, Garching, Germany}
\def   \ie   {{\it i.e.},\ }
\def   \eg   {{\it e.g.},\ }
\def   \etal {{\it et al.}}
\newcommand{\beq}{\begin{equation}}
\newcommand{\eeq}{\end{equation}}
\newcommand{\Mdot}{\dot{M}}
\newcommand{\hii}{H~{\sc ii}~}
\newcommand{\kms}{\mbox{ km s$^{-1}$}}
\newcommand{\cms}{\mbox{ cm s$^{-1}$}}
\newcommand{\gcm}{\mbox{ g cm$^{-3}$}}
\newcommand{\gcms}{\mbox{ g cm$^2$ s$^{-2}$}}
\newcommand{\about}{\mbox{$\sim$}}
\newcommand{\cc}{\mbox{ cm$^3$}}
\newcommand{\Mo}{\mbox{M$_{\odot}$}}
\newcommand{\Lo}{\mbox{$L_{\odot}$}}

\begin{resumen}
Presentamos una breve descripci\'on de la evoluci\'on de la presi\'on del gas 
en regiones de formaci\'on estelar, desde que se forma la nube materna hasta 
que los vientos de las estrellas reci\'en formadas presurizan la regi\'on. Se 
describen los procesos que destruyen la nube y la forma en que se auto-limita 
el n\'umero total de estrellas reci\'en formadas. La alta eficiencia de la
formaci\'on estelar en brotes nucleares es debida a las altas presiones del
gas. Tambi\'en se describe brevemente la evoluci\'on de vientos lentos y 
masivos en regiones de alta presi\'on.

\end{resumen}

\begin{abstract}
We present the main features in the evolution of the gas pressure in star
forming regions, from the formation of the parental cloud to the moment when 
the region is pressurized by interacting stellar winds. The main processes 
for cloud destruction and the self-limiting properties of star formation are 
described. The high star forming efficiency in nuclear starbursts is a 
consequence of the high gas pressures. The evolution of slow winds in
highly pressurized region is also sketched.

\end{abstract}

\section{Introduction}

Galaxies are open systems and the properties of the interstellar gas
are regulated by the internal sources of energy and interactions with
other galaxies. There are a series of different mechanisms, for both
isolated and interacting galaxies, that are able to accumulate large
gas masses and create star-forming clouds in relatively short time
scales. These include cloud collisions, gravitational and thermal
instabilities, Parker instabilities, gas flows in a bar potential,
tidal interactions, direct galaxy collisions, and mergers (some of
these mechanisms are discussed in this volume by Boeker \etal, Borne
\etal, Dultzin-Hacyan, Elmegreen, Friedli \& Martinet, Lamb \etal, and
Moss \& Whittle). Any one, or a combination, of these processes could
be operative in different locations and at different moments in the
host galaxy, and star formation is the end product of a series of
successive condensations of the interstellar medium. Once a cloud is
formed, however, a distinction should be made between molecular and
self-gravitating clouds (see Franco \& Cox 1986). The criterion to
form molecular clouds is high opacity in the UV, to prevent molecule
photo-destruction, and this is achieved at column densities above
$N_{\tau} \sim 5\times 10^{20} (Z/Z_{\odot})^{-1}$ cm$^{-2}$, where
$Z$ is the metallicity and $Z_{\odot}$ is the solar value. In
contrast, self-gravity becomes dominant when the column density
becomes larger than $N_{sg} \sim 5\times 10^{20} (P/P_{\odot})^{1/2}$
cm$^{-2}$, where $P$ is the interstellar pressure and $P_{\odot}$ is
the value at the solar circle. These two values are similar at the
location of the Sun, but $N_{\tau} < N_{sg}$ in the inner Galaxy and
$N_{\tau} > N_{sg}$ in the outer parts of the Milky Way. This
difference has important consequences and may explain the observed
radial trends of molecular gas in spirals: it is easier to form
molecular clouds in the internal, chemically evolved, parts of spiral
galaxies. In any case, the transformation of gas into stars is due to
a gravitational collapse and self-gravity defines the structure of the
star forming clouds.

\section{Self-gravity}

The formation of stellar groups (or isolated stars, if any) occurs in
the densest regions, the cores, of massive and self-gravitating
clouds. In our Galaxy, the $average$ densities for giant molecular
cloud complexes is in the range $10^2$ to $10^3$ cm$^{-3}$, but the
actual densities in the dense cores is several orders of magnitude
above these values: close to about $\sim 10^6$ cm$^{-3}$ (\eg Bergin
\etal\ 1996; see recent review by Walmsley 1995).  Moreover, recent
studies of young stellar objects suggest the existence of even denser
gas, with values in excess of $10^8$ cm$^{-3}$ (Akeson \etal\ 1996).
Thus, the parental clouds have complex clumpy (and filamentary)
structures, with clump-interclump density ratios of about $\sim 10^2$,
or more, and temperatures ranging between 10 and $10^2$ K. In
addition, the existence of large non-thermal velocities, of several km
s$^{-1}$, and strong magnetic fields, ranging from tens of $\mu$G to
tens of mG (see Myers \& Goodman 1988 and references therein),
indicate large $total$ internal pressures, up to more than five orders
of magnitude above the ISM pressure at the solar neighborhood (which
is about $10^{-12}$ dyn cm$^{-2}$). A simple estimate for isothermal,
spherically symmetric, clouds (with a central core of constant density
$\rho_c$ and radius $r_c$, and an external diffuse envelope with a
density stratification $\rho = \rho_c (r/r_c)^{-2}$), indicates that
self-gravity provides these large total pressure values (see
Garc\'{\i}a-Segura \& Franco 1996). In hydrostatic equilibrium, the
pressure difference between two positions located at radii $r_1$ and
$r_2$ from the center of the core is given by $\Delta P =
-\int^{r_2}_{r_1} \rho g_r dr$, where $g_r$ is the gravitational
acceleration in the radial direction. The total pressure at the core
center is
\beq
P(0)=P_0= \frac{2 \pi G}{3} \rho_c^2 r_c^2 + P(r_c)= \frac{8}{5} P(r_c)
\simeq 2\times 10^{-7} \  n_6^2 r_{0.1}^2 \ \ \ \ {\rm dyn \ cm^{-2}},
\eeq
where $G$ is the gravitational constant, $P(r_c)$ is the pressure at
the core boundary $r=r_c$, $n_6=n_c/10^6$ cm$^{-3}$, and
$r_{0.1}=r_c/0.1$ pc. The corresponding core mass is
\beq M_c\simeq \left( \frac{\pi P_0}{G}\right)^{1/2} r_c^2 \sim 10^2 \
P_7^{1/2} r_{0.1}^2 \ \ \ \ {\rm M_{\odot}},
\eeq
where $P_7= P_0/10^{-7}$ dyn cm$^{-2}$. For $P_7\sim 1$ and a typical
core size for galactic clouds, $r_{0.1}\sim 1$ (see Walmley 1995),
gives a value similar to the observationally derived core masses; in
the range of 10 to 300 M$_{\odot}$ (\eg Snell \etal\ 1993). The
pressure inside the core varies less than a factor of two between the
center and $r=r_c$. Taking $r_{0.1}=1$ and the maximum core density
value, $n_c \sim 5\times 10^6$ cm$^{-3}$ (\eg Bergin \etal\ 1996), the
upper bound for the expected core pressures is about $P_0 \simeq
5\times 10^{-6}$ dyn cm$^{-2}$. The large range in observed cloud
properties obviously results in pressure fluctuations of a few orders
of magnitude (both, from cloud to cloud and inside any given cloud),
and it is meaningless to define an ``average'' cloud pressure
value. Actually, given that star forming clouds have nested
structures, in which dense fragments are embedded in more diffuse
envelopes, different cloud locations have different total
pressures. Also, the expected range of cloud pressures in our Galaxy
should probably span from the ISM values, $P_7\sim 10^{-5}$, at the
very external cloud layers, up to $P_7\sim 10$ inside the most massive
star forming cores.

\section{Stellar radiation: HII regions and cloud destruction}

The initial structure and pressure of the gas in a star forming cloud
is defined by self-gravity. Once young stars appear, the new energy
input modifies the structure and evolution of the cloud. Low-mass
stars provide a small energy rate and affect only small volumes,
but their collective action may provide partial support against the
collapse of their parental clouds, and could regulate some aspects of
the cloud evolution (Norman \& Silk 1980; Franco \& Cox 1983; Franco
1984; McKee 1989; see also the paper by Bertoldi \& McKee in this
volume). In contrast, the strong radiation fields and fast stellar
winds from massive stars are able to excite large gas masses and can
even disrupt their parental clouds (\eg Whitworth 1979; Franco \etal\
1994). Also, they are probably responsible for both stimulating and
shutting off the star formation process at different scales. The
combined effects of supernovae, stellar winds, and H II region
expansion destroy star-forming clouds and can produce, at some
distance and later in time, the conditions for further star formation
(\eg Franco \& Shore 1984; Palous \etal\ 1995). Thus, the
transformation of gas into stars may be a self-limited and
self-stimulated process (see reviews by Franco 1991, Ferrini 1992, and
Shore \& Ferrini 1994).
 
In the case of the dense star-forming cores, the sizes of either HII
regions or wind-driven bubbles are severely reduced by the large
ambient pressure (Garc\'{\i}a-Segura \& Franco 1996). In fact, the
pressure equilibrium radii of ultra-compact HII regions are actually
indistinguishable from those of ultra-compact wind-driven bubbles. When
pressure equilibrium is reached, the UCHII radius is
\beq 
R_{{\rm UCHII,eq}} \approx 2.9 \times 10^{-2} \,\, F_{48}^{1/3}
\,\, T_{{\rm HII},4}^{2/3} \,\, P_7^{-2/3} \ \  \,\,{\rm pc} , 
\label{Rsequnits} 
\eeq
where $F_{48}$ is the total number of ionizing photons per unit time in units
of $10^{48}$ s$^{-1}$, and $T_{{\rm HII},4} = T / 10^4$ K. For the case of a
strong wind evolving in a high-density molecular cloud core, the equilibrium 
radius of a radiative bubble is
\beq 
R_{,{\rm WDB,eq}}=\left[ \frac{\dot{M} \,v_{\infty}}{4 \,\pi \,P_0} \right]^{1/2}
\simeq 2.3\times 10^{-2} \left[ \frac{\dot{M_6} \,\,v_{\infty,8}}{P_7}
\right]^{1/2}  \ \  \,\,{\rm pc} , \label{Req} 
\eeq
where the mass loss rate is $\dot{M_6}=\dot{M}/10^{-6}$ M$_\odot$ yr$^{-1}$, 
and the wind velocity is $v_{\infty,8}= v_{\infty}/10^8$ cm s$^{-1}$. Thus, 
for dense cores with $r_c\sim 0.1$ pc, the resulting UCHIIs and wind-driven
bubbles can reach pressure equilibrium without breaking out of the core (\ie 
they could be stable and long lived). Recently, Xie \etal\ (1996) have found 
evidence indicating that this is probably the case: the smaller UCHII seem
to be embedded in the higher pressure cores.

If the limit to continued star forming activity inside the core is due
to photoionization by these internal H II regions, the maximum number
of OB stars is given by the number of H II regions required to
completely ionize the core (Franco et al 1994),
$N_{OB}=(1-\epsilon)M_{c}/M_{i}$, where $M_{i}$ is the ionized
mass. This means that the maximum number of massive stars that can be
formed within a core is
\begin{equation}
N_{OB} \approx 3 {M_{c,2} n_{6}^{3/7} \over F_{48}^{5/7} (c_{s,15}  
t_{MS,7})^{6/7}}
\end{equation}
where $M_{c,2}$ is the core mass in $10^{2}$ M$_{\odot}$, $c_{s,15}$
is the HII region sound speed in units of 15 km s$^{-1}$, and
$t_{MS,7}$ is the mean OB star main sequence lifetime in
$10^{7}$yr. Clearly, for increasing core densities, the value of
$R_{0}$ decreases and the resulting number of OB stars increases. In
the case of the gas in nuclear regions, due to the intrinsic larger
ISM pressures in the inner regions of galaxies, the population of
clouds are denser and more compact. The corresponding star forming
clouds should also be denser than in the rest of the disk, and a
larger number of stars can be formed per unit mass of gas. Thus, {\it
nuclear starbursts can be a natural consequence of the higher pressure
values} (a bursting star formation mode can also be associated to a
delayed energy input, see Parravano 1996).

When stars are located near the edge of the core, and depending on the
slope of the external density distribution, both HII regions and
wind-driven bubbles can accelerate and flare out with a variety of
hydro-dynamical phenomena. These include supersonic outflows, internal
shocks, receding ionization fronts, fragmentation of the thin shell,
etc (\eg Tenorio-Tagle 1982; Franco \etal\ 1989, 1990;
Garc\'{\i}a-Segura \& Mac Low 1995a,b). Thus, no static solution
exists in this case and the pressure difference between the HII
regions and the ambient medium begins to evaporate gas from the
cloud. This represents a clear and simple physical mechanism for cloud
destruction and, as the number of OB stars increases, more expanding H
II regions form and limit the rate of new star formation by ionizing
the surrounding molecular gas (Franco et al 1994).  Eventually, when
the whole cloud is completely ionized, star formation ceases.  The
total cloud mass ionized by an average OB star, integrated over its
main sequence lifetime, is
\begin{equation}
M_{i}(t) \approx \frac{2\pi}{3} R_{0}^{3}\mu_p n_{0}
\left\lbrack\left(1+\frac{5c_{s}t_{MS}}{2R_{0}}\right)^{6/5}-1\right\rbrack.
\end{equation}
where $R_{0}$ is the initial radius at the average cloud density,
$n_{0}$, $\mu_p$ is the mass per gas particle, $c_{s}$ is the sound
speed in the HII region, and $t_{MS}$ is the main sequence lifetime of
the average OB star. For a cloud of mass $M_{GMC}$, with only 10\% of
this mass in star-forming dense cores, the number of newly formed OB
stars required to completely destroy it is
\begin{equation}
N_{OB} \sim 30 
\frac{M_{GMC,5}n_{3}^{1/5}}{F_{48}^{3/5}(c_{s,15} t_{MS,7})^{6/5}}.
\end{equation}
where $M_{GMC,5}=M_c/10^5$ \Mo, $n_{3}=n_{0}/10^3$ cm$^{-3}$,
$c_{s,15}=c_{s}/15$ km s$^{-1}$, and $t_{MS,7}=t_{MS}/10^7$ yr.
Assuming a standard IMF, this corresponds to a total star forming
efficiency of about $\sim 5$ \%. For the average values of stellar
ionization rates and giant molecular cloud parameters in our Galaxy,
the overall star forming efficiency should be about 5\%. Obviously,
larger average densities and cloud masses can result in higher star
formation efficiencies.

Summarizing, photoionization from OB stars can destroy the parental
cloud in relatively short time scales, and defines the limiting number
of newly formed stars. The fastest and most effective destruction
mechanism is due to peripheral, blister, HII regions, and they can
limit the star forming efficiency at galactic scales. Internal HII
regions at high cloud pressures, on the other hand, result in large
star forming efficiencies and they may be the main limiting mechanism
in star forming bursts and at early galactic evolutionary stages (see
Cox 1983).

\section{Mechanical energy}

As the cloud is dispersed, the average gas density decreases and the
newly formed cluster becomes visible. The individual HII regions merge
into a single photo-ionized structure and the whole cluster now powers
an extended, low density, HII region. The stellar wind bubbles now can
grow to larger sizes and some of them begin to interact. As more winds
collide, the region gets pressurized by interacting winds and the
general structure of the gas in the cluster is now defined by this
mass and energy input (Franco \etal\ 1996).

Given a total number of massive stars in the cluster, $N_{OB}$, and their 
average mass input rate, $<\dot{M}>$, the pressure due to interacting adiabatic
winds is
\begin{equation}
P_i\sim \frac{N_{OB} <\dot{M}> c_i}{4 \pi r^2_{clus}}\sim 10^{-8}
\frac{N_{2} <\dot{M}_6> c_{2000}}{r^2_{pc}} \ \ \ \ {\rm dyn \ cm^{-2}},
\end{equation}
where $N_{2}=N_{OB}/10^2$, $<\dot{M}_6>=<\dot{M}>/10^{-6}$ M$_\odot$
yr$^{-1}$, $r_{pc}=r_{clus}/1$ pc is the stellar group radius, and
$c_{2000}=c_i/2000$ km s$^{-1}$ is the sound speed in the interacting
wind region. This is the central pressure driving the expansion of the
resulting superbubble before the supernova explosion stage. For modest
stellar groups with relatively extended sizes, like most OB
associations in our Galaxy, the resulting pressure is only slightly
above the ISM pressure (\ie for $N_{2} \sim 0.5$ and $r_{pc}\sim 20$,
the value is $P_i \sim 10^{-11}$ dyn cm$^{-2}$).  For the case of rich
and compact groups, as those generated in a starburst, the pressures
can reach very large values. For instance, for the approximate cluster
properties in starbursts described by Ho in this volume, $r_{pc}\sim
3$ and $N_{2}>10$, the resulting pressures can reach values of the
order of $P_1\sim 10^{-7}$ dyn cm$^{-2}$, similar to those due to
self-gravity in star forming cores. At these high pressures, the winds at the
evolved red giant (or supergiant) phases cannot expand much and they reach 
pressure equilibrium at relatively small distances from the evolving star. 
Thus, the large mass ejected during the slow red giant 
wind phase is concentrated in a dense circumstellar shell.
\begin{figure}
\vspace*{43mm}
\begin{minipage}{43mm}
\includegraphics{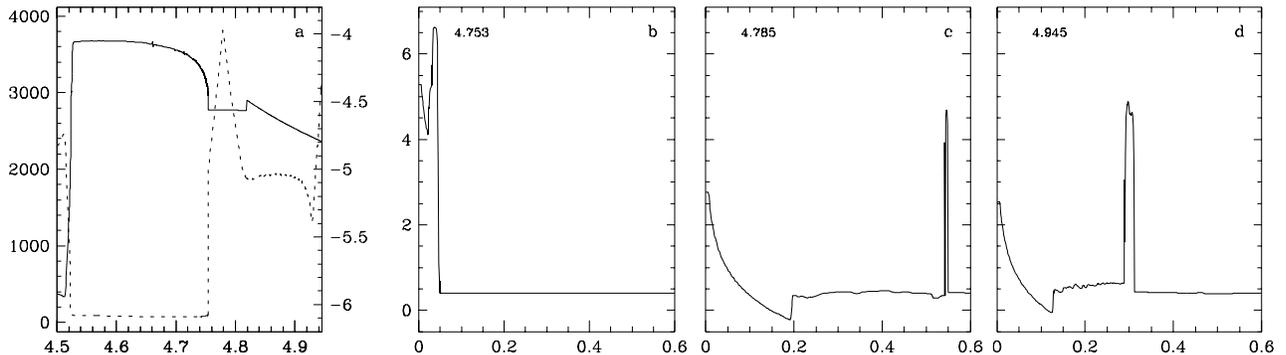}
\end{minipage}

\caption{
Evolution of a wind-driven bubble in a high pressure medium.
{\bf (a)} Evolution of a stellar wind from a 35\,M$_{\odot}$
star. Left scale: terminal wind velocity (km s$^{-1}$); right scale:
mass loss rate (M$_{\odot}$ yr$^{-1}$); horizontal scale: time
(millions of yr). {\bf (b)-(d)}: Evolution of the wind-driven bubble. The gas
density (cm$^{-3}$) is plotted in a logarithmic scale against the radial 
distance (pc). Evolutionary times (shown in the upper-left corner of each 
panel) are given in million years.}
\label{fig:1}
\end{figure}
An example of this is shown in Figure 1, where the evolution of a
wind-driven bubble around a 35\,M$_{\odot}$ star is presented. Fig. 1a
shows the wind velocity and mass-loss rate (dashed and solid lines,
respectively: Garcia-Segura, Langer \& Mac Low 1996). We ran the
simulation only over the time spanning the red supergiant and
Wolf-Rayet phases, and assume that the region is already pressurized
by the main sequence winds from massive stars.  We used the AMRA code,
as described by Plewa \& R\'o\.zyczka in this volume.  During the RSG
phase the wind-driven shell is located very close ($R\sim$ 0.04 pc) to
the star due to a very low wind ram-pressure (Fig. 1b). Later on
(Fig. 1c), the powerful WR wind pushes the shell away from the star to
the maximum distance of $R\approx 0.54$ pc. Still later, when the wind
has variations, the shell adjusts its position accordingly, and
reaches the distance $R\sim 0.3$ pc at the end of simulation
(Fig. 1d). It must be stressed that the series of successive
accelerations and decelerations of the shell motion during the WR
phase will certainly drive flow instabilities and cause deviations
from the sphericity assumed in our model. The role of these
multidimensional instabilities in the evolution of the shell is
currently under study (with 2-D and 3-D models), and the results will
be presented in a future communication.

Regardless of the possible shell fragmentation, however, when the star
explodes as a supernova, the ejecta will collide with a dense
circumstellar shell. This interaction generates a bright and compact
supernova remnant, with a powerful photoionizing emission (\ie
Terlevich \etal\ 1992; Franco \etal\ 1993; Plewa \& R\'o\.zyczka this
volume), that may also be a very strong radio source, like SN 1993J
(see Marcaide \etal 1995). If the shell is fragmented, the
ejecta-fragment interactions will occur during a series of different
time intervals, leading to a natural variability in the emission at
almost any wavelength (see Cid \etal 1996). This type of interaction
is also currently under investigation, and further modeling will shed
more ligth on the evolution of SN remnants in high-pressure environs.

This work was done during the First ``{\bf Guillermo Haro}'' Workshop,
in April-May 1996, and we thank the hospitality of the staff at INAOE
in Tonantzintla, Puebla. We warmly thank many useful discussions with
Roberto Cid, Jos\'e Rodriguez-Gaspar, Elisabete de Gouveia, Gustavo
Medina-Tanco, Michal R\'o\.zyczka, Sergei Silich, Laerte Sodre,
Guillermo Tenorio-Tagle, Roberto Terlevich, and To\~ni Varela, and the
enthusiasm and support given to the whole project by Alfonso Serrano,
General Director of INAOE. JF and GGS acknowledge partial support to
this project by DGAPA-UNAM grant IN105894, CONACyT grants
400354-5-4843E and 400354-5-0639PE, and a R\&D Cray research
grant. The work of TP was partially supported by the grant KBN
2P-304-017-07 from the Polish Committee for Scientific Research. The
simulations were carried out on a workstation cluster at the
Max-Planck-Institut f\"ur Astrophysik.


{\small
\begin{references}

\reference 
Akeson, R. L., Carlstrom, J. E., Phillips, J. A. \& Woody, D. P. 1996,
ApJL, in press

\reference  
Bergin, E., Snell, R. \& Goldsmith, P. 1996, ApJ, in press

\reference 
Cid Fernandes R., Plewa T., R\'o\.zyczka M., Franco J., Tenorio-Tagle G., 
Terlevich R. \& Miller W., 1996, MNRAS, in press

\reference
Cox, D. P. 1983, ApJL, 265, L61

\reference
Ferrini, F. 1992, in {\it Evolution of Interstellar Matter and Dynamics of
Galaxies}, ed. J. Palous, W. B. Burton \& P. O. Lindblad, (Cambridge: 
Cambridge Univ. Press), 304

\reference
Franco, J. 1984, A\&A, 137, 85

\reference
Franco, J. 1991, in {\it Chemical and Dynamical Evolution of Galaxies}, ed.
F. Ferrini, J. Franco \& F. Matteucci, (Pisa: ETS Editrice), 506

\reference
Franco, J. \& Cox, D. P. 1983, ApJ, 273, 243

\reference
Franco, J. \& Cox, D. P. 1986, PASP, 98, 1076

\reference
Franco, J., Garc\'{\i}a-Segura, G. \& Plewa, T. 1996, in preparation

\reference
Franco, J., Miller, W., Cox, D., Terlevich, R., R\'o\.zyczka, M.
\& Tenorio-Tagle, G. 1993, RevMexAA, 27, 133

\reference
Franco, J. \& Shore, S. N. 1984, ApJ, 285, 813

\reference
Franco, J., Shore, S. N. \& Tenorio-Tagle, G. 1994, ApJ, 436, 795

\reference
Franco, J., Tenorio-Tagle, G. \& Bodenheimer, P. 1989, RMxAA, 18, 65

\reference
Franco, J., Tenorio-Tagle, G. \& Bodenheimer, P. 1990, ApJ, 349, 126

\reference
Garc\'{\i}a-Segura, G. \& Franco, J. 1996, ApJ, in press

\reference
Garc\'{\i}a-Segura, G., Langer, N., \& Mac Low, M.-M. 1996, A\&A, in press

\reference
Garc\'{\i}a-Segura, G. \& Mac Low, M.-M. 1995a, ApJ, 455, 145

\reference
Garc\'{\i}a-Segura, G. \& Mac Low, M.-M. 1995b, ApJ, 455, 160

\reference
Marcaide, J. M., Alberdi, A., Ros, E., \etal 1995, Science, 270, 1475

\reference
McKee, C. F. 1989, ApJ, 345, 782

\reference
Myers, P. C. \& Goodman, A. A. 1988, ApJ, 326, L27

\reference
Norman, C. \& Silk, J. 1980, ApJ, 238, 158

\reference
Palous , J., Tenorio-Tagle, G. \& Franco, J. 1994, MNRAS, 270, 75

\reference
Parravano, A. 1996, ApJ, 462, 594.

\reference
Snell, R., Carpenter, J., Schloerrb, F. P. \& Strutskie, M. 1993, in 
{\it Massive Stars: Their Lives in the Interstellar Medium}, ed. J. P. 
Cassinelli \& E. B. Churchwell, ASP (Conf. Series) 35, 138

\reference
Shore, S. N. \& Ferrini, F. 1994, FundCosmicPhys, 16, 1

\reference
Tenorio-Tagle, G. 1982, in {\it ``Regions of Recent Star Formation''}, ed. R. 
S. Roger \& P. E. Dewdney, (Dordrecht; Reidel)

\reference
Terlevich, R., Tenorio-Tagle, G., Franco, J. \& Melnick, J., 1992
MNRAS, 255, 713

\reference 
Xie, T., Mundy, L. G., Vogel, S. N. \& Hofner, P. 1996, ApJ, submitted

\reference 
Walmsley, M. 1995, RevMexAA (Conf. Ser.), 1, 137

\reference 
Whitworth, A. 1979, MNRAS, 186, 59

\end{references}
}
\end{document}